%% file: isca14Final.tex
\documentclass[final]{jpaper}
\usepackage[normalem]{ulem}

\placetextbox{.285}{0.085}{\footnotesize{978-1-4799-4394-4/14/\$31.00~\copyright~2014~IEEE}}

\usepackage{epstopdf}
\usepackage{epsfig}
\usepackage{amssymb} 
\usepackage[tbtags]{amsmath} 
\usepackage{graphics,eepic,epic}
\usepackage{latexsym}
\usepackage{euscript}
\usepackage{tabularx}
\usepackage{subfig}
\usepackage{graphics,eepic,epic,psfrag}
\usepackage{color}
\usepackage{algorithm, algorithmic}

\newcommand{\pProfile}{SleepScale}
\newcommand{\specialcell}[2][c]{%
  \begin{tabular}[#1]{@{}l@{}}#2\end{tabular}}
  \newcommand{\yanpei}[2][c]{%
  \begin{tabular}[#1]{@{}c@{}}#2\end{tabular}}

\newcommand{\yp}[1]{{\color{red} [#1]}}

\newcommand{\bbE}{\mathbb{E}}

\newtheorem{theorem}{Theorem}

\usepackage{cite}

\begin{document}

\title{SleepScale: Runtime Joint Speed Scaling and Sleep States Management \\ for Power
Efficient Data Centers}
\author{
Yanpei Liu$^{\dagger}$\quad~Stark C. Draper$^{\star}$\quad~Nam Sung Kim$^{\dagger}$ \\
$^{\dagger}$University of Wisconsin Madison \quad $^{\star}$University of Toronto\\
\authemail{yliu73@wisc.edu  \quad stark.draper@utoronto.ca \quad nskim3@wisc.edu}\\
}

\date{}

\maketitle

\thispagestyle{empty}

\begin{abstract}
Power consumption in data centers has been growing significantly in
recent years. To reduce power, servers are being equipped with
increasingly sophisticated power management mechanisms. Different
mechanisms offer dramatically different trade-offs between power
savings and performance penalties. Considering the complexity,
variety, and temporally-varying nature of the applications hosted in a
typical data center, intelligently determining which power management
policy to use and when is a complicated task.

In this paper we analyze a system model featuring both performance scaling and low-power states. We reveal the interplay between performance scaling and low-power states via intensive simulation and analytic verification. Based on the observations, we present \pProfile, a runtime power management tool designed to efficiently exploit existing power control mechanisms. At run time, \pProfile~characterizes power consumption and quality-of-service
(QoS) for each low-power state and frequency
setting, and selects the best policy for a given QoS constraint. We evaluate \pProfile~using workload traces from data centers and achieve significant power savings relative to conventional power management strategies.
\end{abstract}
\section{Introduction}
\label{sec.intro}
\input{introduction}
\section{System Model}
\label{sec.background}
\input{systemModel}


\section{Workload-dependent Optimal Policy}
\label{sec.singleserver}
\input{analysis}


\section{\pProfile}
\label{sec.policyprofiling}
\input{numResults}

\section{Evaluation}
\label{sec.evaluation}
\input{evaluation}

\section{Conclusion and Future Work}
\label{sec.conclusion}
\input{conclusion}

\section*{Acknowledgment}
\input{ack}

\appendix
\label{sec.appendix}
\input{theorems}

\bstctlcite{bstctl:etal, bstctl:nodash, bstctl:simpurl}
\bibliographystyle{IEEEtranS}
\bibliography{myrefs}

\end{document}

%% file: introduction.tex
As the workloads allocated to data centers increase so does the
economic and environmental footprint of these processing
clusters. U.S. data center electricity consumption grew by roughly
$36\%$ between $2005$ and $2010$ to about $2\%$ of domestic usage in
$2010$ (this is about $77,728,000~MW \cdot hr/year$, or $12$ billion
U.S. dollars) \cite{Koomey}. Further, power usage by U.S. data centers
is doubling every five years \cite{AmurCipar}. These numbers give
material economic, societal, and environmental reasons for
improving power efficiency.  

Much power consumed by
data centers does not go toward computation. While configured to peak service demand, data centers regularly operate at lighter load levels, and are often only
$20-30\%$ utilized \cite{BarrosoHolzle}.  Even in idle mode a server
continues to draw roughly $60\%$ of the power of a busy
server \cite{xeon, BarrosoHolzle}.  Although some data
centers (e.g.~in Google) recently report low power usage effective (PUE),
benefiting from the cooling using cold water directly from rivers or
seas, around $40\%$ of costs are still associated with power
consumption in other commercial and governmental
data centers \cite{CostofCloud}. Thus, it is important for data
centers to maximize their power efficiency and at the same time to guarantee a certain level of QoS.

To maximize power efficiency, commercial processors and platforms
support various power management mechanisms such as dynamic
voltage/frequency scaling (DVFS) and low-power (sleep)
states. However, it is the power management policies (i.e., the choice
of operating voltage/frequency and the determination of when to enter
which low-power state) that ultimately govern power efficiency for
given workload under target QoS constraints. Thus in this paper, the word ``policy'' refers to some combination of power control methods such as processing speed and low-power state settings.
 

To develop effective policies we consider a system model that
takes into account important QoS constraints and  processor,
platform, and workload characteristics
(Section~\ref{sec.background}). 
Then, to develop engineering insight, we study the effectiveness of
various policies for the aforementioned system model under idealized setting (Section~\ref{sec.singleserver}). In contrast to some previous
studies, we demonstrate that there is not a single optimal policy; the
optimum policy heavily depends on QoS constraints, as well as on the
characteristics of processor, platform, and workload. Through intensive simulation and analytic study,
we observe that ($1$) there exists an
optimal joint choice of frequency setting and low-power state; ($2$)
the best low-power state depends on the performance constraint at
low utilization; ($3$) the best low-power state also depends on the job size; ($4$) the delay to enter a low-power state should be jointly
determined with frequency and ($5$) service time dependency on CPU frequency matters.

Based on our analyses and engineering insights, we develop a power management tool, {\em SleepScale}, that
selects the best combination of frequency and low-power states at runtime. It
predicts the characteristics of given workload and determines the optimal
power management policy with low overhead. While we note that accurately predicting the
workload characteristics of a computing system (i.e., the utilization, service time and inter-arrival time distributions) is the key to effective runtime power management, we also demonstrate
that simple prediction techniques can offer sufficient accuracy for real-world
utilization traces.

The paper is outlined as follows. We summarize related work in
Section~\ref{sec.related_work}. Power and operational models are
introduced in
Section~\ref{sec.background}. Section~\ref{sec.singleserver} presents
engineering lessons for an idealized workload. \pProfile~is
introduced in Section~\ref{sec.policyprofiling} and evaluated in
Section~\ref{sec.evaluation}. We conclude in
Section~\ref{sec.conclusion}.

\input{relatedwork}

%% file: relatedwork.tex
\section{Background and Related Work}
\label{sec.related_work}


Modern computer systems are equipped with two classes of power
management mechanisms to reduce power consumption: performance scaling
and component deactivation.  DVFS is a widely used performance scaling
technique where one reduces processing speed and voltage in step with
reduced utilization.  In component deactivation one puts the CPU and
other peripherals into a low-power sleep
state. Table~\ref{tb.cpupowerstate} shows typical power states for the
Intel Xeon/Atom family of processors.  We now summarize some related work on both types of mechanisms.

\begin{table}[!t]
\small
	\renewcommand{\arraystretch}{1.1}
	\centering
	\begin{tabular}{ll}
	\toprule
	{\bf State} & {\bf Operation \& Characteristics}\\
	\midrule
	$C0_{(a)}$ & Operating active state: there is work to do, voltage \& \\
& frequency setting adjusted dynamically by DVFS\\
	$C0_{(i)}$ & Operating idle state: there is no work to do, voltage \& \\
& frequency held constant at last DVFS setting\\
	$C1$ & Halt state: clock stops\\
	$C3$ & Sleep state: cache flushed, architectural state maintained, \\
 & clock stopped\\
	$C6$ & Deep sleep state: architectural state saved to RAM, \\& voltage set to zero  \\
	\bottomrule
	\end{tabular} \\ 
		\caption{CPU power states \cite{xeon}.}
	\label{tb.cpupowerstate}
\end{table}

{\bf Power management via performance scaling.} Performance scaling
such as DVFS has been widely used for computer systems to provide
substantial power savings by varying voltage and frequency
\cite{Herbet, Kaxiras, Snowdon}. In \cite{GeorgeHarrison} the problem
of optimal speed scaling is formulated as a stochastic dynamic problem
and a numerical solution is derived. In \cite{Andrew} the speed
scaling problem is studied via an algorithmic approach and also
jointly considered with scheduling. Recently, in \cite{MemScale},
performance scaling in memory systems is considered. In~\cite{CoScale}
methods are developed to scale the speed of memory systems and
processors in a joint manner.

{\bf Power management via sleep states.} Various low-power states are
designed for halting processors when they are in idle. In
\cite{MeisnerGold} the authors propose a method for eliminating idle
power in servers by quickly transitioning between a high-performance
active state and a single low-power state. The development is based on
queuing theory \cite{mor_book}. Recent advances in the $M/G/k$ queue
with setup costs \cite{GandhiHarcholAdan} extend the case to the
multi-server scenario: the impact of data center size is studied in
\cite{GandhiHarchol} and power allocation in server farms is studied
in \cite{GandhiHarcholDas}. In a slightly different vein \cite{Neely}
takes a stochastic optimization approach, optimizing time average
system performance using optimization theory. However due to distinct
wake-up penalties, determining when to enter what low-power state is a
complicated task. In \cite{MedanBuyuktosunoglu} the authors warn of
potential problems of using these low-power states and suggest
``guarded'' mechanisms to avoid negative power savings.

{\bf Joint approaches and \pProfile.}
Recently, it is suggested that halting when the system is idle and using a static rate when busy perform almost as good as an optimal speed scaling mechanism \cite{Wierman}. However the low-power state model in~\cite{Wierman} is rather limited and only one single low-power state is considered. In \cite{MeisnerGold} the DVFS mechanism is considered in separation from sleep states. Other approaches have also been proposed, for instance a real-time server probing mechanism is proposed in \cite{Tang1}. A more static approach based on workload profiling is proposed in \cite{Tang2}. 

Nevertheless, much is still not known about which sleep
state/speeds are useful when, and how decisions should be made with low overhead
to toggle between power saving policies. \pProfile~takes a unified approach, modeling a general system
with speed scaling and many low-power states having different
characteristics.  It seeks to develop a family of policies that
jointly manage the setting of the operating frequency and the choice of which sleep state to enter, and more importantly, how these decisions should be made online with low overhead. 

%% file: systemModel.tex

We now present a system model that accounts for both DVFS and
low-power states. We later use the model to study 
performance and power consumption under various workloads.

\subsection{Power model}

\begin{table*}[t]
\small
	\renewcommand{\arraystretch}{1.1}
	\centering
	\begin{tabular} {cccccc}
	\toprule
	{\bf Components} & {\bf Operating} & {\bf Idle} & {\bf Sleep} & {\bf Deep sleep} & {\bf Deeper Sleep} \\ 
	\midrule
	CPU$\times 1$ \cite{xeon} & $130V^2 f~W$ ($C0_{(a)}$) & $75V^2 f~W$ ($C0_{(i)}$) & $47V^2~W$ ($C1$) & $22~W$ ($C3$) & $15~W$ ($C6$) \\
	Chipset$\times 1$ \cite{xeon} & $7.8~W$ & $7.8~W$ & $7.8~W$ & $7.8~W$ & $7.8~W$ \\
	RAM$\times 6$ \cite{xeon} & $23.1~W$ & $10.4~W$ & $10.4~W$ & $10.4~W$ & $3.0~W$\\ 
	HDD$\times 1$ \cite{hdd} & $6.2~W$ & $4.6~W$ & $4.6~W$ & $4.6~W$ & $0.8~W$ \\
	NIC$\times 1$ \cite{nic} & $2.9~W$ & $1.7~W$ & $1.7~W$ & $1.7~W$ & $0.5~W$ \\
	
	Fan$\times 1$ \cite{MeisnerGold} & $10~W$ & $1~W$ & $1~W$ & $1~W$ & $0~W$ \\
	
	PSU$\times 1$ \cite{MeisnerGold} & $70~W$ & $35~W$ & $35~W$ & $35~W$ & $1~W$ \\
	
	Platform total & $120~W$ ($S0_{(a)}$) & $60.5~W$ ($S0_{(i)}$) & $60.5~W$ ($S0_{(i)}$) & $60.5~W$ ($S0_{(i)}$) & $13.1~W$ ($S3$) \\
	\bottomrule
	\end{tabular} \\ 
		\caption{Power consumption for different components of a system.}	
	\label{tb.powernumber}
\end{table*}

\begin{table}[t]
\small
	\renewcommand{\arraystretch}{1.1}
	\centering
	\begin{tabular}{lll}
	\toprule
	{\bf State} & {\bf Operation \& supported CPU state(s)}\\
	\midrule
	$S0_{(a)}$ & Active state: associated with $C0_{(a)}$ only \\
        $S0_{(i)}$ &  Idle state: associated with other CPU states \\
	$S3$ & Sleep: RAM powered, associated with $C6$ only \\
	
	\bottomrule
	\end{tabular} \\ 
		\caption{Platform power states \cite{platformState}.}
	\label{tb.platformpowerstate}
\end{table}




We discuss how we model (i) the power consumed by the processor,
(ii) the power consumed by peripheral (non-CPU) components such as
DRAM, hard disk drive (HDD), network interface card (NIC), and (iii)
the latencies involved in transitioning between power states.  

First,
consider processor power.  A CPU in the active $C0_{(a)}$ state (and
also in the idling $C0_{(i)}$ state) consumes dynamic power.  Power
consumption will be proportional to $ V^2 f $, where $V$ is the supply
voltage and $f \in [0,1]$ is the DVFS clock frequency scaling
factor. In our study, we choose a linear DVFS scenario, where both
voltage and frequency are scaled linearly. This assumption falls
within the scope of some existing processors (see the datasheet in
\cite{previousData}). Dynamic power consumption in states $C0_{(a)}$
and $C0_{(i)}$ will therefore scale cubically with frequency.  In the
sequel we consider only the frequency parameter $f$ and assume $V$ to
be proportional to $f$. In sleep state $C1$ the clock signal is gated.
Thus, only leakage power is consumed. Platform components also have
low-power states and each supports a subset of the CPU states.
Table~\ref{tb.platformpowerstate} lists the platform power states and
the CPU state (or states) that each supports.

The power consumption of the entire system is the sum of CPU power
and platform power.  In the following we use the term ``state'' to
encompass both CPU and platform state and denote the combined state
by concatenating their notations, e.g., $C0_{(i)}S0_{(i)}$.
Table~\ref{tb.powernumber} tabulates power consumption numbers for the Xeon
family of CPUs and associated platform components.
As an example, the power consumption in state $C0_{(i)}S0_{(i)}$ is $75V^2f
+ 52.7~W$.


Table~\ref{tb.wakeuplatency} summarizes the delay incurred by the
various possible states in returning to active operation.  We note
that while it is possible to consider a platform shut-down in which
the entire system is turned off, the wake-up latency incurred will be
enormous, and thus should be considered at a coarser time granularity
than is the focus of this work.

\begin{table}[t]  
\small
	\renewcommand{\arraystretch}{1.1}
  \centering
    \begin{tabular}{cccccc}
  \toprule
	 & $C0_{(a)}$ & $C0_{(i)}$ & $C1$ & $C3$ & $C6$ \\ \midrule
    $S0_{(a)}$ & $0~s$   & --  & -- & -- & --  \\ 
    $S0_{(i)}$ & --   & $0~s$   & $1-10~\mu s$ & $10 - 100~\mu s$ & $0.1 - 1~ms$  \\ 
    $S3$ & -- & --   & -- & -- & $1 - 10~s$  \\ \bottomrule
  \end{tabular} \\
  	\caption{Average wake-up latencies\cite{MeisnerGold, GuevaraLubin}.}	
	\label{tb.wakeuplatency}
\end{table}

\subsection{Operation model}
\label{sec.system_model}

We assume jobs arrive at the system according to some random process with rate $\lambda$ and are served based on the first-come-first-serve (FCFS) order. The server
is equipped with a DVFS mechanism, which affect the service time.  In
active operation the clock frequency can be scaled by a factor $f \in
[0, 1]$ and the time it takes to process each job is scaled
correspondingly. For CPU-bound jobs, the resulting (scaled) service
times are modeled as a random variable with mean $\frac{1}{\mu f}$ where $\mu$ is the max service rate. Setting
the frequency to the maximum $f = 1$ yields maximum processing speed
$\frac{1}{\mu}$ and setting $f = 0$ stops the server from processing jobs,
i.e., the server is in a clock-gated mode. For memory-bound jobs,
  the service time is modeled as independent of frequency, thus with
  mean $\frac{1}{\mu}$. Finally, we note that the ``utilization'' factor
$\rho = \frac{\lambda}{\mu}$ is the expected fraction of time the server has
jobs to process.

As discussed earlier, in  active state $C0_{(a)}S0_{(a)}$ power
varies cubically in $f$; hence the power is $P_0 f^3$ for some maximum
power $P_0$. The system also has $n$ low-power states indexed by $i$,
$1 \leq i \leq n$.  Each time the server becomes idle it enters a
sequence of low-power states, staying in each for a
pre-set amount of time. The server enters state $i$ some $\tau_i$
seconds after its queue empties.  Naturally, $\tau_1 < \tau_2 < \ldots
< \tau_n$.  Formally, the $i$th low-power state is characterized by
the three-tuple $(P_i, \tau_i, w_i)$ where:
\begin{itemize}
\item  $P_i$ is the power consumed in state $i$, 
\item $\tau_i$ is the time at which the server enters state $i$,
measured from the time the queue empties, and
\item $w_i$ is the average wake-up latency that the system incurs to
  return to the active state $C0_{(a)}S0_{(a)}$ from state $i$.
\end{itemize}

A job arrival interrupts the low-power state and wakes up the system
from its current low-power state. We assume a job arrival immediately
triggers the wake-up process, during which no job can be
served. Deeper sleep states consume less power but take longer to wake
up from so $P_1 > P_2 > \ldots P_n$ but $w_1 < w_2 < \ldots < w_n$.
For simplicity and conservative evaluation we assume the power
consumption during the wake-up is the same as the power consumed in
active operation.

The low-power states that we study include $C0_{(i)}S0_{(i)}$,
$C1S0_{(i)}$, $C3S0_{(i)}$, $C6S0_{(i)}$ and $C6S3$. We also analyze
sequences of pairs of low-power states such as entering
$C0_{(i)}S0_{(i)}$ first then $C6S3$, and also sequences many more
low-power states, e.g., entering $C0_{(i)}S0_{(i)}$, $C1S0_{(i)}$,
$C3S0_{(i)}$, $C6S0_{(i)}$ and $C6S3$ in sequence.

%% file: analysis.tex
In this section we study the model of Section~\ref{sec.background} under the
idealized assumptions of Poisson arrivals and exponentially
distributed service times. We present our methodology for evaluating
various policy choices and discuss engineering lessons.

\subsection{Methodology}
\label{sec.methodology}

To evaluate a candidate policy we generate $N = 10,000$ jobs and
evaluate the policy at each possible frequency setting based on our
model presented in Section~\ref{sec.background}. The simulated maximum
frequency is $f = 1$ and the minimum is the one that the system is
barely stable, i.e., $f = \rho + 0.01$ with step size of $0.01$. (We
take such a fine step size only to generate smooth plots, in a real system
there would be about 10 distinct frequencies.)  For policies that
consist of a sequence of low-power states, when the processor queue
empties the processor enters the low-power states one after another.
The arrival of the next job wakes up the processor, incurring some
latency. From our simulations we gather results on mean response time
$\bbE[R]$ and average power $\bbE[P]$. An example of our simulator is
provided in Algorithm~\ref{alg.simulator} for single low-power state
with $\tau_1 = 0$. Simulating one policy, i.e., one frequency and
low-power state combination takes on average $6.3~ms$ on an Intel i5
$2.6$ GHz machine using Matlab. This can take even less time when an
optimized code is dedicated to run Algorithm~\ref{alg.simulator}.

\begin{algorithm}[!t]
\caption{Simulation under $\rho = \frac{\lambda}{ \mu}$ and frequency $f$}
\label{alg.simulator}
\begin{algorithmic}[1]
\STATE Generate jobs with job size and inter-arrival time sampled from probability distributions with mean $\frac{1}{\mu f}$ and $\frac{1}{\lambda}$ respectively (assuming CPU-bound).
\FOR {job $j$ in $1$ to $N$}
\IF {job $j$ arrives before $j-1$ departs}
\STATE Active += service time of $j$.
\STATE Delay of $j$ = departure time of $j$ - arrival time of $j$.
\ELSE
\STATE Active += service time of $j$ + wake-up latency. 
\STATE Idle += arrival time of $j$ - departure time of $j-1$.
\STATE Delay of $j$ = service time of $j$ + wake-up latency.
\ENDIF
\ENDFOR
\STATE Compute delay by taking average of all $j$ jobs.
\STATE Compute power by the ratio of active and idle periods. 
\end{algorithmic}
\end{algorithm}

\subsection{Engineering lessons}
\label{sec.lessons}

\begin{figure}
\centering
\subfloat[DNS-like: $\rho = 0.1$]{\label{fig.DNS_ER_rho01} \includegraphics[width=0.45\textwidth]{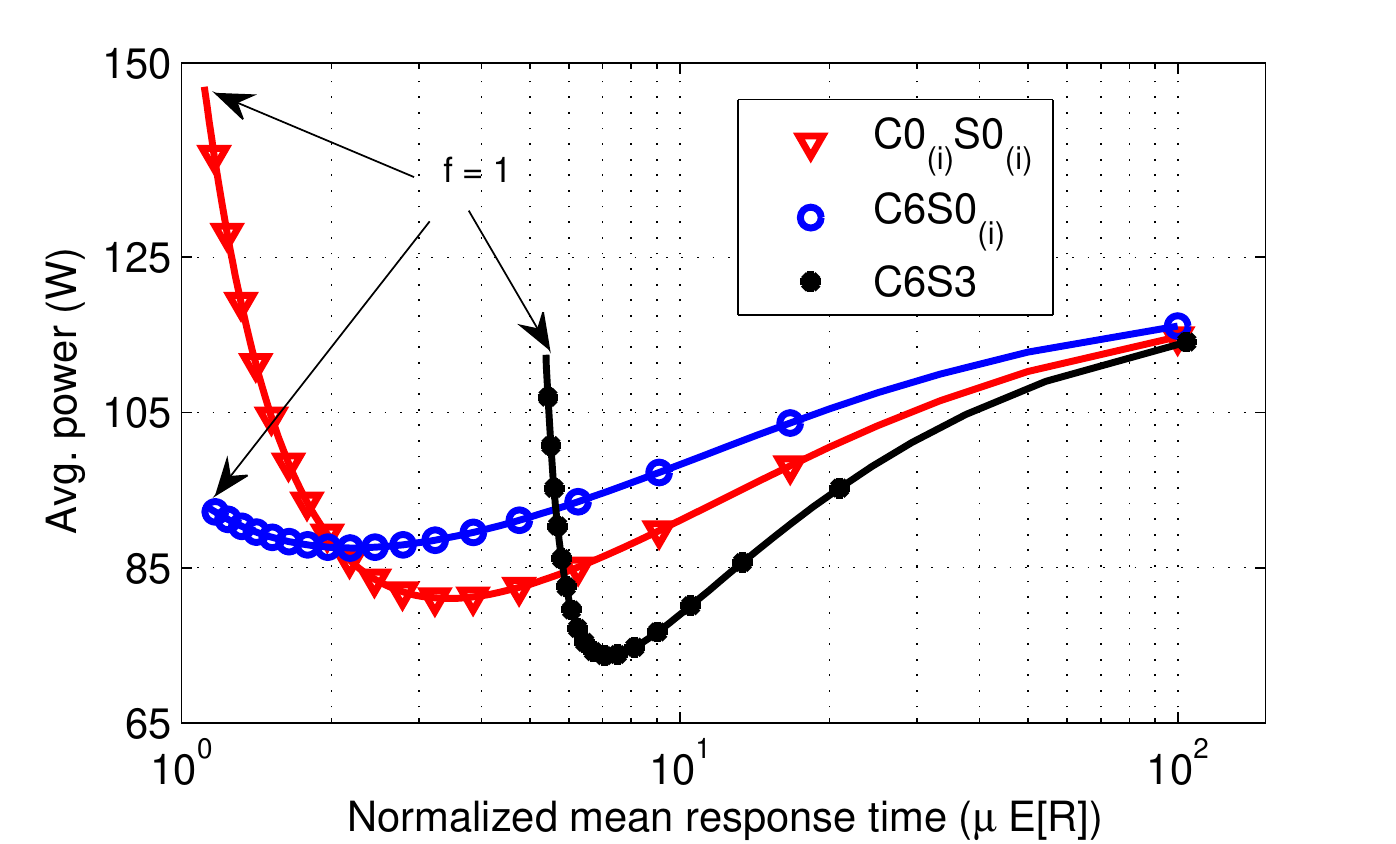}} \\
\vspace*{-1 em}
\subfloat[Google-like: $\rho = 0.1$]{\label{fig.Google_ER_rho01} \includegraphics[width=0.45\textwidth]{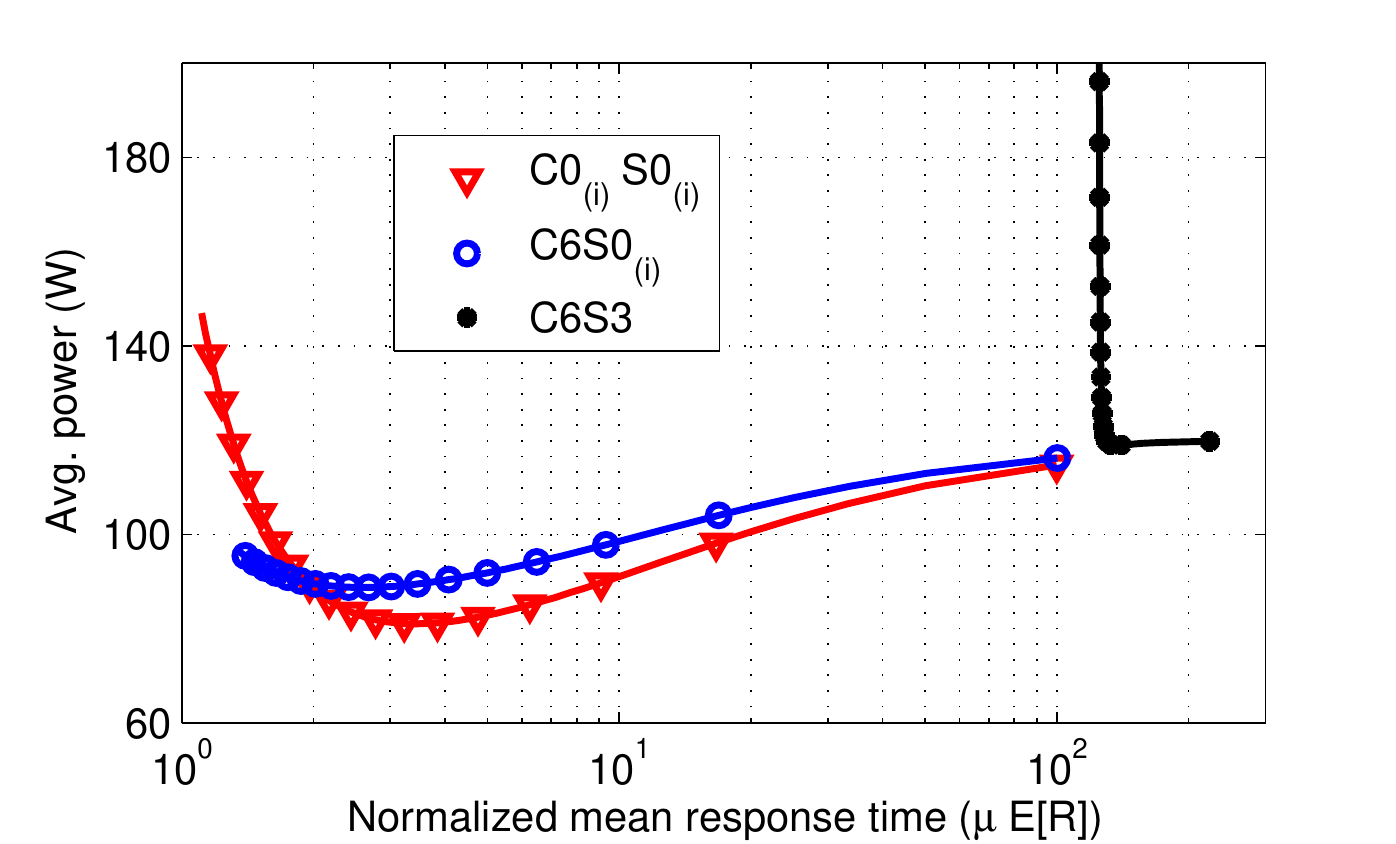}} 
\caption{Mean response time $\bbE[R]$ and average power $\bbE[P]$ trade-off. Only representative low-power states are shown for the clarity of illustration.}
\label{fig.ER_EP}
\end{figure}


We first consider $\tau_1 = 0$, i.e., whenever the server completes all jobs in its queue the
server immediately enters a low-power state from the active state
$C0_{(a)}S0_{(a)}$.  Second, to investigate how job size effect the
choice of policy, we consider two different sizes: ``Google-like''
($\frac{1}{\mu} = 4.2~ms$) and ``DNS-like'' ($\frac{1}{\mu} = 194~ms$). In
Section~\ref{sec.policyManager}, when we use traces from data centers
for our analysis, the former is roughly the size of a Google web
search job and the latter of a DNS look-up job
(cf.~Table~\ref{tb.applications}). Unless otherwise specified, jobs
are assumed to be CPU-bound (and thus service time scales linearly in
frequency).


The average wake-up latencies for the system to return to the active
$C0_{(a)}S0_{(a)}$ state from various low-power states are set as
follows.  To wake-up from $C1S0_{(i)}$ we set the average latency to
$10~\mu s$, from $C3S0_{(i)}$ we set it to $100~\mu s$, from
$C6S0_{(i)}$ we set it to $1~ms$, and from $C6S3$ we set it to $1~s$.
These values fall in the ranges specified in
Table~\ref{tb.wakeuplatency}.  {\em We observe that other choices from
the range specified do not greatly change the following engineering
lessons.}

We begin our engineering lesson by introducing
Figure~\ref{fig.ER_EP}. In each sub-figure we plot average power
consumption $\bbE[P]$ as a function of two parameters.  The first
parameter is the mean response time $\bbE[R]$ normalized by the
average service time $\frac{1}{\mu}$.  We normalize the mean response time in
order to compare different workloads (with different average job
sizes) directly.  The second parameter is the DVFS frequency setting
$f$.  Selected choices of $f$ are indicated by the hash marks on each
curve. As $f$ is changed the effective service rate $\mu f$ varies
proportionally.  The left end of each curve is $f=1$, corresponding to
the fastest processing, and hence smallest response time but most
power. The right end of the curve is set by the smallest frequency
under which the system is stable, roughly $f = \rho$, where $\rho$ is
the utilization. The hash marks are spaced uniformly in increments of
$0.05$.

We also evaluate systems based on Atom processors with the power
numbers from \cite{GuevaraLubin}. While simulation results will not be
shown, we will discuss our observations.

{\bf $1$) There exists an optimal joint choice of frequency setting
and low-power state.} Figure~\ref{fig.ER_EP} plots the DNS-like and
Google-like workloads results for $\rho = 0.1$.  We observe that there
is an optimal joint choice of frequency setting and low-power state
that yields the minimum power consumption. Without a constraint on the
mean response time, the optimal choice corresponds to the bottom of
the lowest bowl.  For example, in Figure~\ref{fig.ER_EP}-(a) the
globally optimum policy uses $C6S3$ and $f=0.42$ and runs at an
average power of $70~W$. Setting $f$ too high leads to worsening
efficiency since power increases cubically in frequency.  But, setting
it too small means that each job takes longer to complete, thereby
reducing the possibility of entering a low-power state. The optimal
choice -- the policy at the bottom of the bowl -- strikes a balance
between these two effects. Policies such as
``race-to-halt'' \cite{MeisnerGold} wherein the server runs at maximum
frequency $f = 1$ until the queue empties and then immediately enters
a single low-power state can consume $50\%$ more power. Such a policy
corresponds to the leftmost tip of each curve.

Due to small processor power and relatively large platform power, for
Atom processors running DNS-like jobs at low utilizations, it is
better to run fast and enter low-power state immediately after the job queue empties. 


{\bf $2$) At low utilization, the best low-power state depends on the
mean response time budget.} In Figures~\ref{fig.ER_EP} the left-hand
side of the plots corresponds to a tight requirement on normalized
mean response time.  As the constraint is loosened different choices
of policies and operating frequency settings become optimal.  For
example, in Figure~\ref{fig.ER_EP}-(a) under the tightest constraint
(when the normalized mean response time $\mu \bbE[R]$ is required to
be in the range $[1,2]$) policies using $C6S0_{(i)}$ are optimal;
under a mid-range constraint policies using $C0_{(i)}S0_{(i)}$ are the
best; and under the loosest constraint policies using $C6S3$ prove to
be the best. The intuition is as follows.

Consider the range of normalized mean response time for which
$C6S0_{(i)}$ is the best in Figure~\ref{fig.ER_EP}-(a).  This is the
best choice for the tightest response time requirements because
setting the frequency high results in faster job completion, thereby
increasing the opportunities to enter a more aggressive power saving
mode such as $C6$. On the other hand, under a looser constraint on
response time the frequency can be lowered yielding the cubic decrease
in power consumption.  In this situation the expected duration of an
empty queue is small and thus it becomes too costly to enter states
like $C6$ that have high wake-up latencies.  Thus $C0_{(i)}S0_{(i)}$
outperforms $C6S0_{(i)}$ in this situation. This observation is also valid for systems of Atom processors running
Google-like workload.


{\bf $3$) The best power state depends on the job size.} 
\begin{figure}
\centering
\includegraphics[width = 0.45\textwidth]{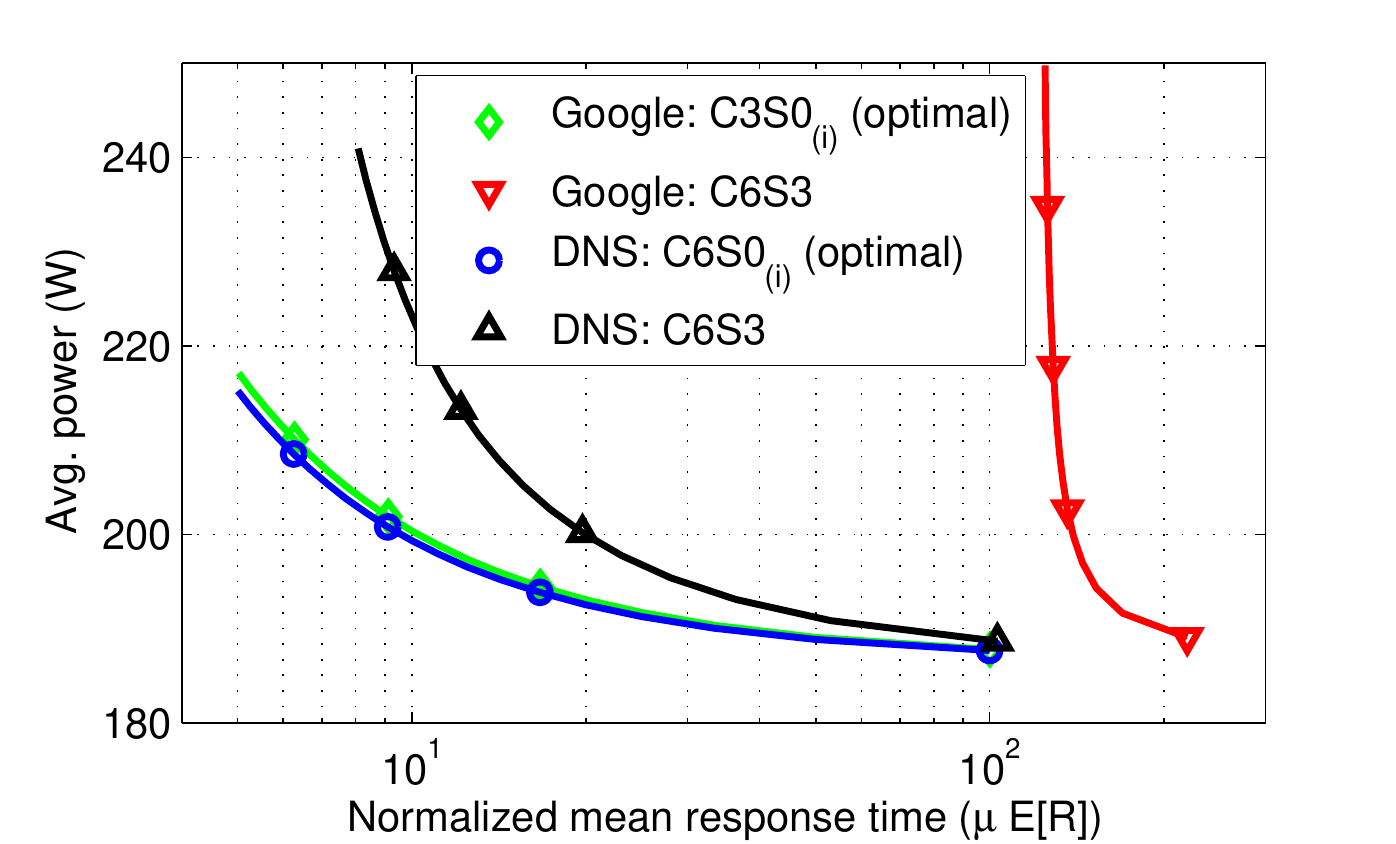}
\caption{Optimal low-power states for Google and DNS-like workload under high utilization. Other sub-optimal low-power states are not shown for the clarity purpose.}
\label{fig.ER_rho08_DNSGoogle}
\end{figure}
Figure~\ref{fig.ER_rho08_DNSGoogle} shows the optimal low-power states
among other possible ones (not all shown) for
servers under high utilization.  When heavily utilized the
server rarely exits the active operating state
$C0_{(a)}S0_{(a)}$. This reduces opportunities to realize power
savings by transitioning to a sleep state, thus most power savings
must come from DVFS.

However the job size also plays a crucial rule in the optimal
low-power state. For the DNS-like workload, policies using
$C6S0_{(i)}$ dominate since the wake-up latency of $C6S0_{(i)}$ is
negligible compared to average job size ($194~ms$). But, for
Google-like workload the relatively small job size is sensitive to
large wake-up latencies, thus policies using $C3S0_{(i)}$ becomes
optimal. For the same reason, very aggressive sleep states such as
$C6S3$ should not be used for small size jobs or should be used only
during extremely long idle period, ``guarded'' by workload prediction
techniques \cite{MedanBuyuktosunoglu}. Similar observations can also be made in Atom-based systems.




{\bf $4$) The delay $\tau_i$ to enter a low-power state should be
jointly determined with frequency.} When a server becomes idle, it may wait some amount of time before entering a low-power
state to avoid unnecessary wake-up costs~\cite{GandhiHarchol}. Of
course, for some low-power states with very small wake-up latencies,
there is no need to delay the entrance as the wake-up incurs
negligible penalty. However, for other low-power states with heavy
wake-up penalties, it is not immediately clear how long the system
should wait.

\begin{figure}
\centering
\includegraphics[width = 0.45\textwidth]{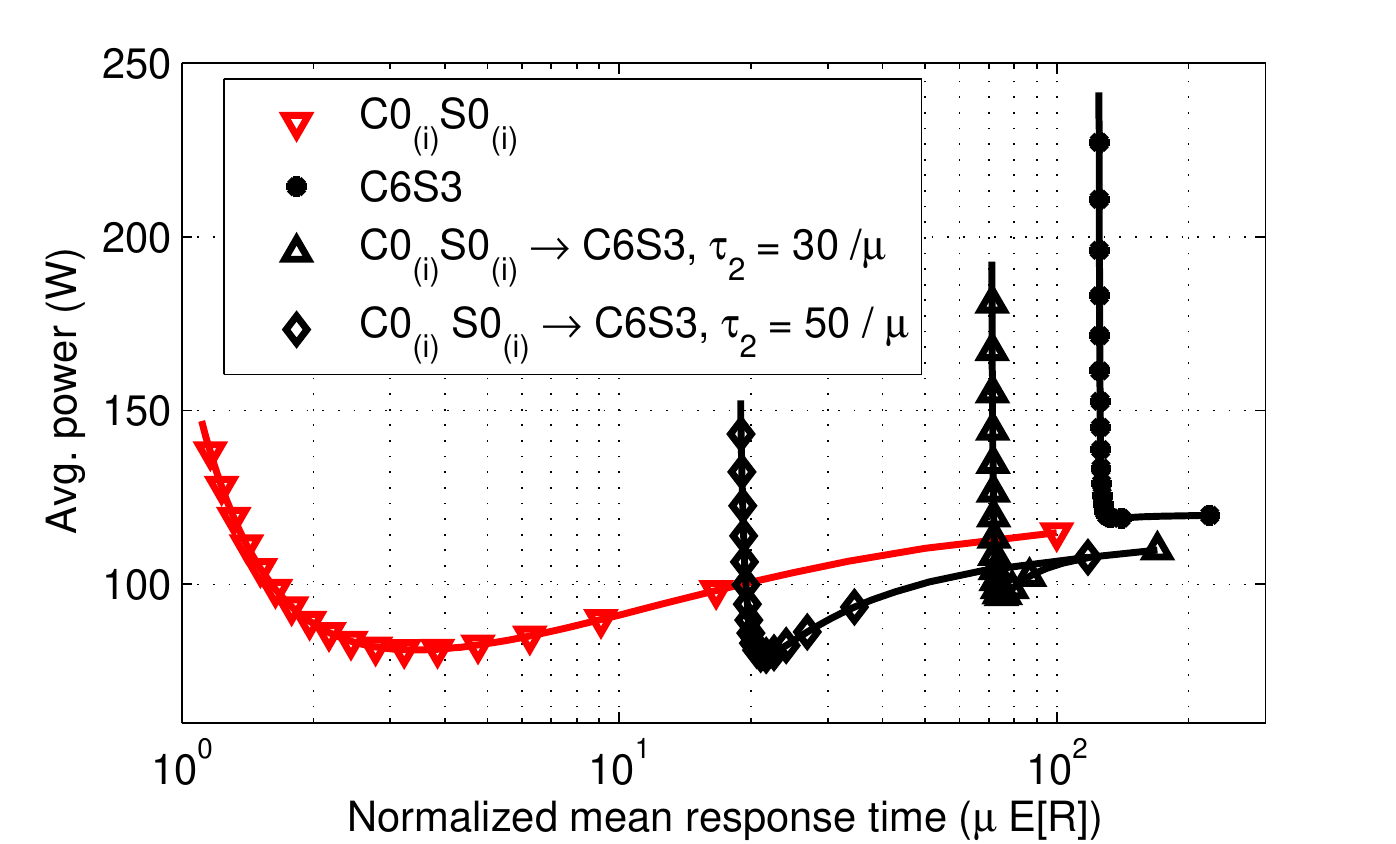}
\caption{Entering the second low-power state after delays for Google-like workload.}
\label{fig.delayed_enter}
\end{figure}

Figures~\ref{fig.delayed_enter} considers this situation and show what
happens if we delay the entrance to state $C6S3$ by various
amounts. In ``$C0_{(i)}S0_{(i)} \rightarrow C6S3$'', the server first
enters $C0_{(i)}S0_{(i)}$ immediately ($\tau_1 = 0$) whenever the job
queue empties and will enter $C6S3$ if it idles for some $\tau_2$
seconds (recall the definition of $\tau_i$ in
Section~\ref{sec.system_model}). We compare policies using delayed
$C6S3$ with ones using immediate $C6S3$ and immediate
$C0_{(i)}S0_{(i)}$. We observe that the delay parameter $\tau_2$
interpolates between the immediate $C6S3$ and $C0_{(i)}S0_{(i)}$
curves: setting $\tau_2 = 0$ reduces to immediate $C6S3$ and setting
$\tau_2 = \infty$ reduces to immediate $C0_{(i)}S0_{(i)}$. From the
plots we observe that by delaying $C6S3$, more power savings can be
made at mild mean response time budget (e.g.~consider $\mu \bbE[R] = 20$).  Essentially, what
we observe is that there is an optimal combination between frequency
and entrance delay that minimizes the power under a certain mean
response time constraint.

{\bf $5$) Sequential power throttle-back is conservative.} It is
tempting to concatenate all low-power states i.e., building a system
with large number of low-power states and then letting the system
enter those states in sequence to derive the most benefit from each.
However from our intensive simulation on entering $C0_{(i)}S0_{(i)}$, $C1S0_{(i)}$, $C3S0_{(i)}$, $C6S0_{(i)}$ and $C6S3$ in sequence,  we discover that such policies
are not often efficient.  The reason is that at high utilization the
system rarely enters the last state.  At low utilization it is a waste
of power not to go to the optimal state immediately. Nevertheless,
such sequential power reducing policies can be useful when the arrival
statistics are unknown.

\begin{figure}
\centering
\includegraphics[width = 0.45\textwidth]{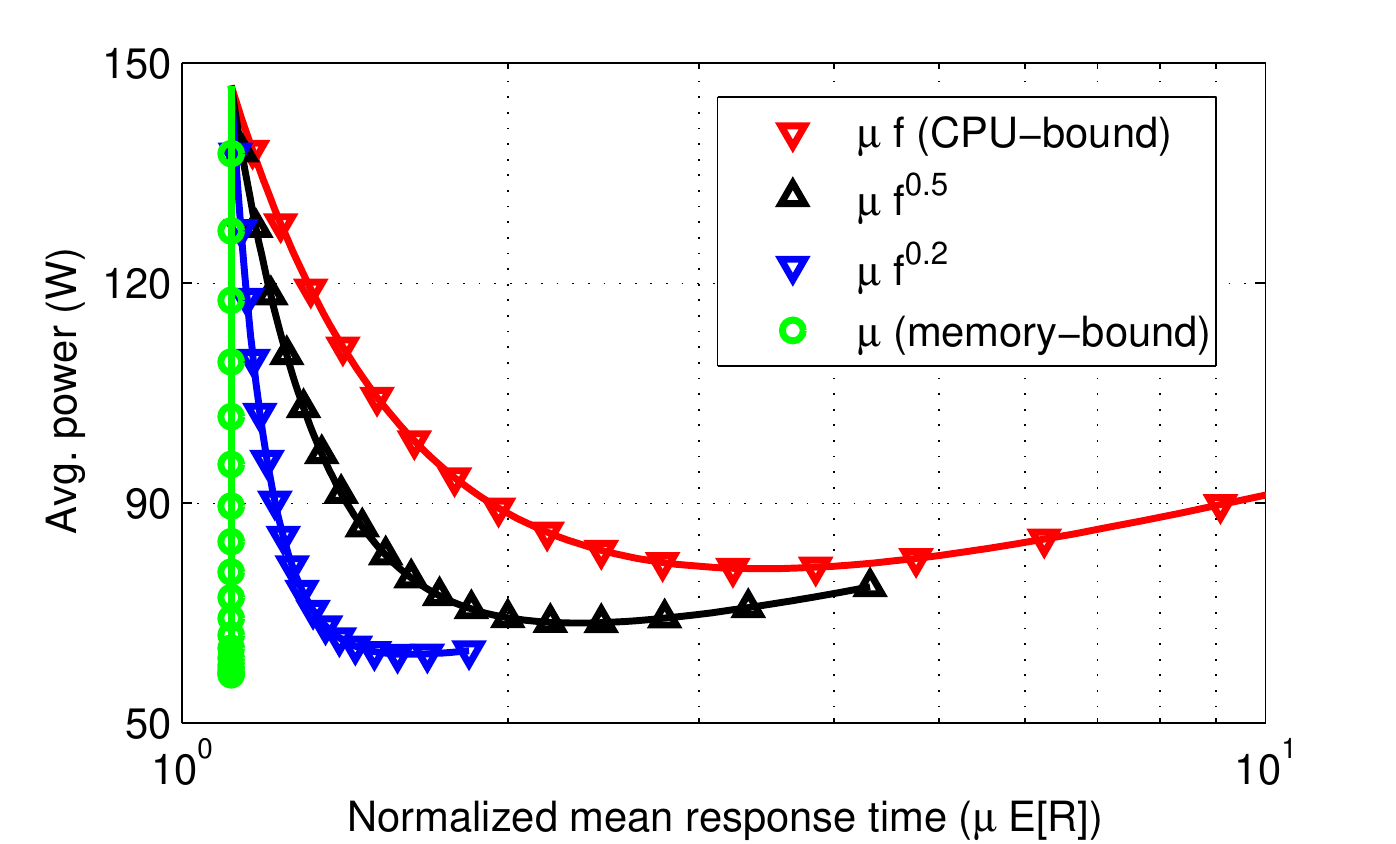}
\caption{Comparison between different CPU usages for DNS-like workload.}
\label{fig.CPU-Mem-bound}
\end{figure}

{\bf $6$) Service time dependency on CPU frequency matters.}  Now we discuss what happens when the workload is {\em not}
CPU-bound. Recall that for CPU-bound jobs the service rate $\mu$
scales linearly with frequency $f$. For less CPU-bound jobs, the
service rate scales sub-linearly and in the extreme case
(memory-bound) service rate is insensitive to frequency (as job
completion time is dominated by memory access time rather than
processing time).  In Figure~\ref{fig.CPU-Mem-bound} we plot a
DNS-like workload at low utilization for service rate varying
as different functions of clock frequency.  It can be observed that
the optimal choice of frequency depends on the scaling. For memory-bound
jobs, the optimal speed is the lowest speed.

\subsection{Analytic  results}
In fact, under the Poisson job arrival process and exponential service time distribution we can derive in closed-form the average power $\bbE[P]$ and
mean response time $\bbE[R]$ for the system model described in
Section~\ref{sec.background}. The results obtained from the closed-form expressions
match those presented in Figure~\ref{fig.ER_EP}. Note that
constructing Figure~\ref{fig.ER_EP} using closed-form expression does
not involve simulations described in Section~\ref{sec.methodology}. The closed-form solution can also be derived when the service time follows general distributions, i.e., not limited to exponential distribution \cite{mor_book}. However for both general arrival process and service time distribution (which SleepScale assumes, as will be discussed next), no closed-form solution exists to the best of our knowledge. We
include these theoretical results in the Appendix for reference.


%% file: numResults.tex
In this section we present \pProfile, a runtime power management
algorithm.  It consists of a policy manager and a runtime predictor.
First, in Section~\ref{sec.policyManager} we detail the
policy manager that selects the optimum policy as a function of
workload statistics.  Second, in Section~\ref{sec.runTimePredictor} we
describe our runtime predictor of the statistical characteristics of
the workload such as utilization, and arrival rate and service time
distributions. This allows \pProfile~to select the best policy in an
online manner.





\subsection{Policy manager}
\label{sec.policyManager}

In this section we describe the policy manager.  The manager takes a
statistical description of the current workload as input and determines
the best policy.




\subsubsection{Policy characterization and selection.}
\label{sec.workloadstat}


The policy manager bases its determination of the best policy on the
empirically observed distributions of recent arrivals and service
times, collected from the server at runtime. The observed recent arrivals and service times can be
arbitrary statistics, not limited to Poisson process and exponential distributions. Given the collected statistics
it characterizes the power-performance trade-off of each low-power
state at a range of frequency settings.  It does this by simulating
the queuing process as described in Algorithm~\ref{alg.simulator}.
The optimal policy is the policy that minimizes power consumption
while meeting a target QoS constraint. As noted in
Section~\ref{sec.methodology}, simulating a single policy takes only
$6~ms$. Considering the finite number of frequencies and low-power
states, the overhead of simulating all policies is thus negligible
compared to the policy updating period (which will be measured in
minutes, as discussed later).





Our QoS constraint is determined by a baseline system.  Our baseline
is motivated by the fact that data centers are often provisioned to
meet a QoS target for some peak demand, often specified in an SLA.  To meet SLA
commitments during periods of peak demand, the data center should be
running full out, i.e., at maximum frequency $f = 1$ without using
a low-power state. In contrast, at lower loads there is slack in meeting the QoS
which can be exploited to reduce operating costs (e.g., power) as much
as possible.

\begin{figure}
\centering
\includegraphics[width = 0.45\textwidth]{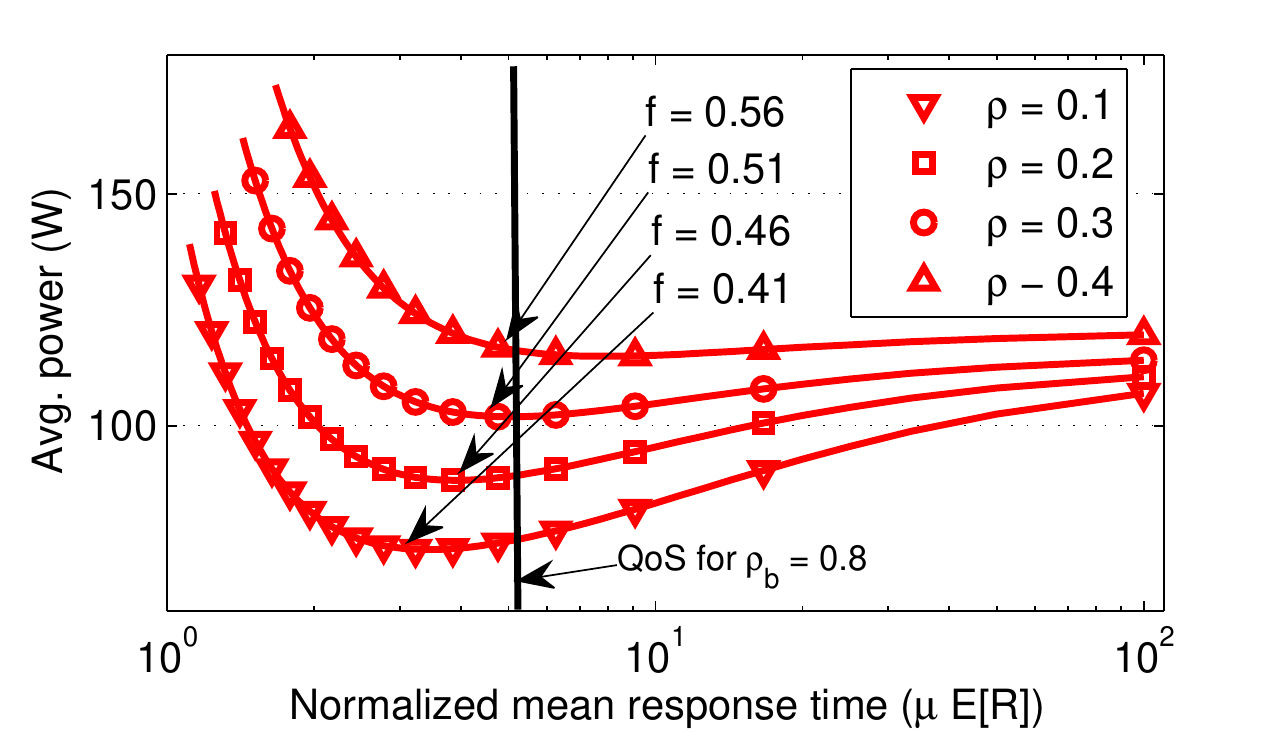}
\caption{Average power/performance trade-off for Google-like workload.}
\label{fig.Google_ER_different_rho}
\end{figure}

We parameterize the target peak demand through a peak design
utilization $\rho_b$.  To understand the baseline QoS, consider
Figure~\ref{fig.Google_ER_different_rho}. This figure plots the
power-delay trade-off for the Google-like workload running with
low-power state $C0_{(i)}S0_{(i)}$ at different frequency settings and
under different utilizations $\rho < \rho_b$. In the plot the
baseline QoS throughput constraint is indicated by the vertical bar.
This vertical bar indicates that the allowable normalized response
time is $5$ when $\rho_b = 0.8$, calculated (under the idealized
model) as $\mu \bbE[R] = \frac{1}{(1-\rho_b)} = \frac{1}{1 - 0.8} =
5$.  In general as the utilization $\rho$ is increased from $\rho = 0$
to $\rho = \rho_b$ the curves shift up, meaning that a higher
frequency setting is required to maintain the QoS throughput
constraint.  For instance, for $\rho =0.4$, which is strictly less
than $\rho_b = 0.8$, to minimize average power one must set $f = 0.56$
and the system will operate exactly at the requisite QoS.  However,
for even lower utilizations, e.g, $\rho = 0.1$, one operates at the
lowest average power by setting $f = 0.41$.  This is the global
minimum for this utilization and the normalized mean response time
achieved is about $3$, which exceeds the QoS requirement.  Thus, one
can sometimes exceed the baseline QoS while minimizing power
consumption.



\begin{table}[!t]
\small
	\renewcommand{\arraystretch}{1.1} \centering \begin{tabular}
	{ccccc} \midrule {\bf Workload} & \yanpei{{\bf
	Inter-arrival} \\ {\bf Mean}} & \yanpei{{\bf Inter-arrival} \\
	$C_v$} & \yanpei{{\bf Service} \\ {\bf Mean}} & \yanpei{{\bf
	Service} \\ $C_v$} \\ \midrule DNS & $1.1~s$ & $1.1$ &
	$194~ms$ & $1.0$ \\ Mail & $206~ms$ & $1.9$ & $92~ms$ &
	$3.6$ \\ Google & $319~\mu s$ & $1.2$ & $4.2~ms$ &
	$1.1$ \\ \bottomrule \end{tabular} \\ \caption{Different
	workload types from \cite{BigHouse} (partial
	list).}  \label{tb.applications}
\end{table}

\begin{figure*}[!t]
\centering
\subfloat[DNS-like with $\bbE{[R]}$ constraint]{\label{fig.PP_DNS_ER} \includegraphics[width=0.5\linewidth]{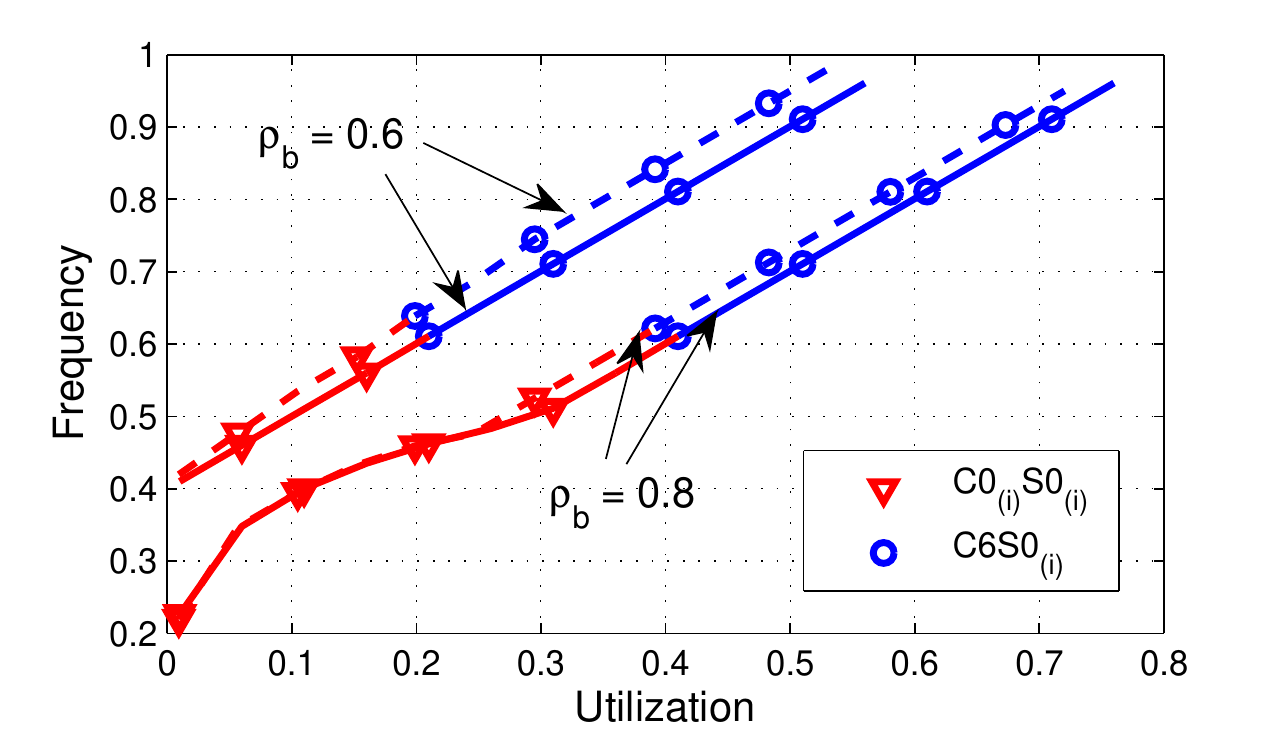}} \hspace*{-1.5em}
\subfloat[Google-like with $\bbE{[R]}$ constraint]{\label{fig.PP_Google_ER} \includegraphics[width=0.5\linewidth]{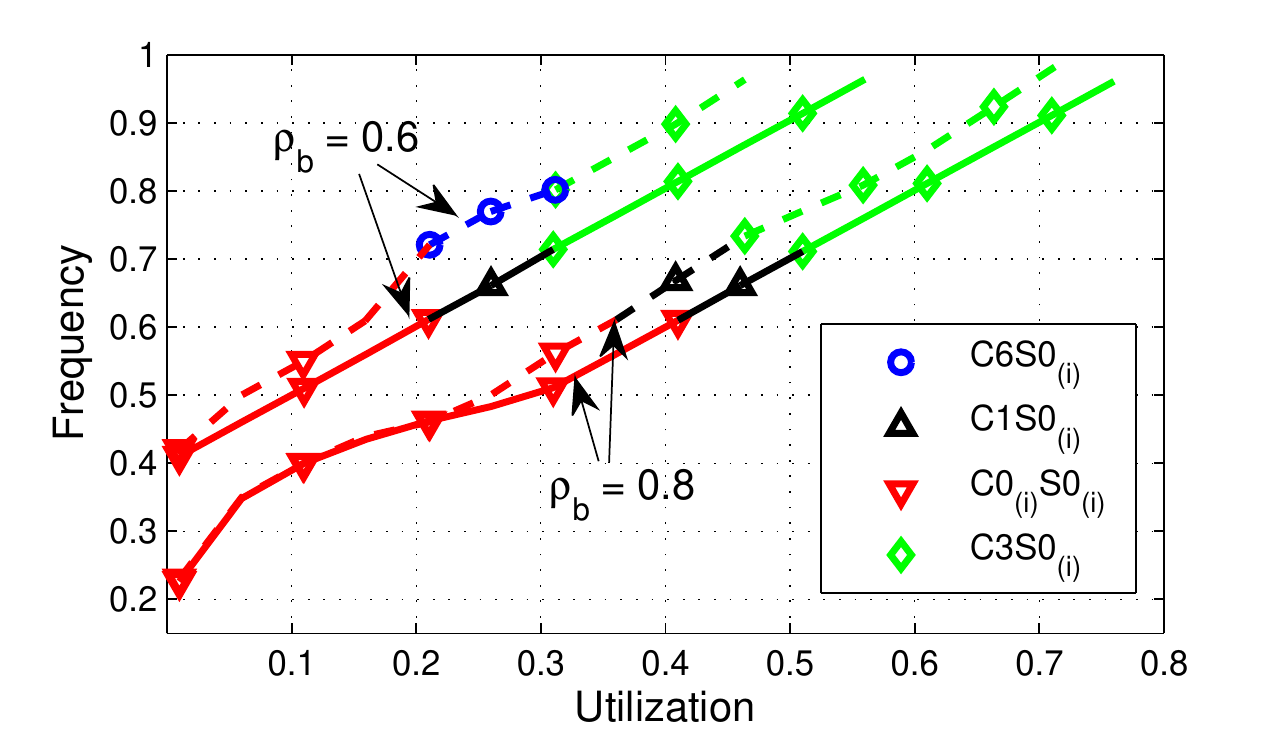}}  
 \\
\vspace*{-1 em}

\subfloat[DNS-like with $Pr(R \geq d)$ constraint]{\label{fig.PP_DNS_Prob_d} \includegraphics[width=0.5\linewidth]{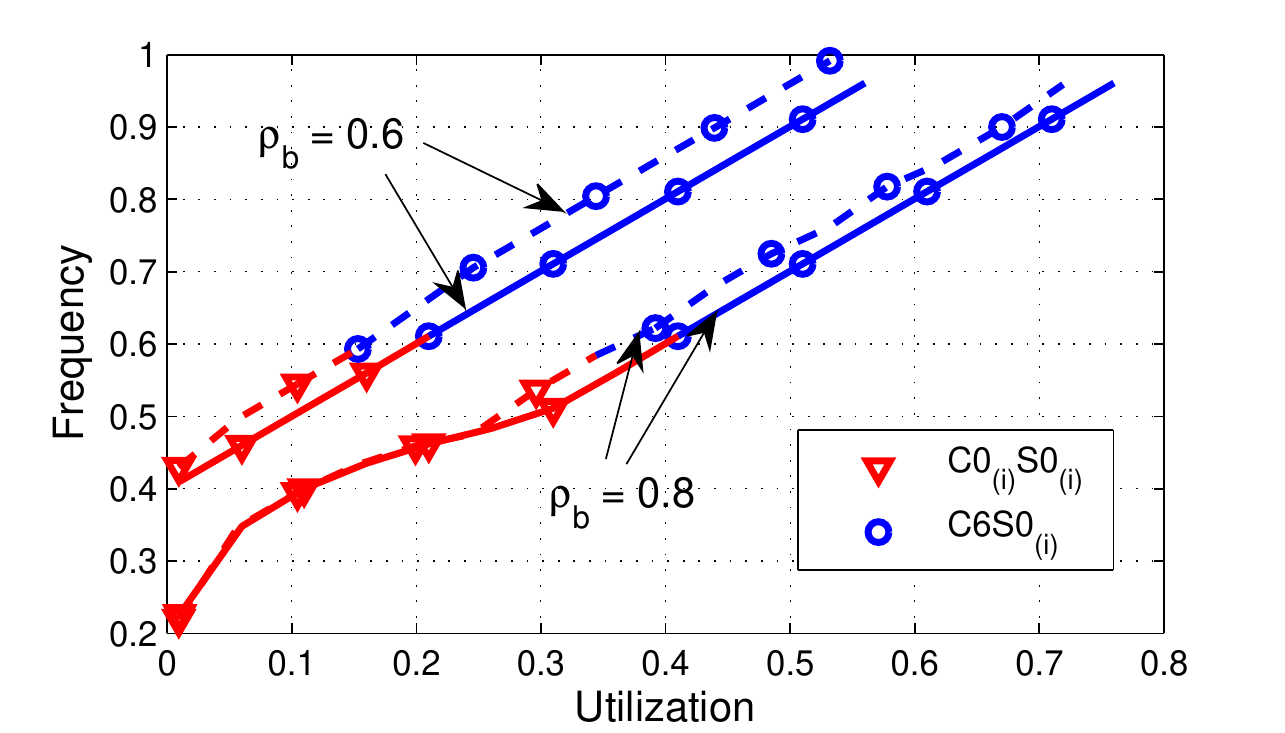}} \hspace*{-1.5em}
\subfloat[Google-like with $Pr(R \geq d)$ constraint]{\label{fig.PP_Google_Prob_d} \includegraphics[width=0.5\linewidth]{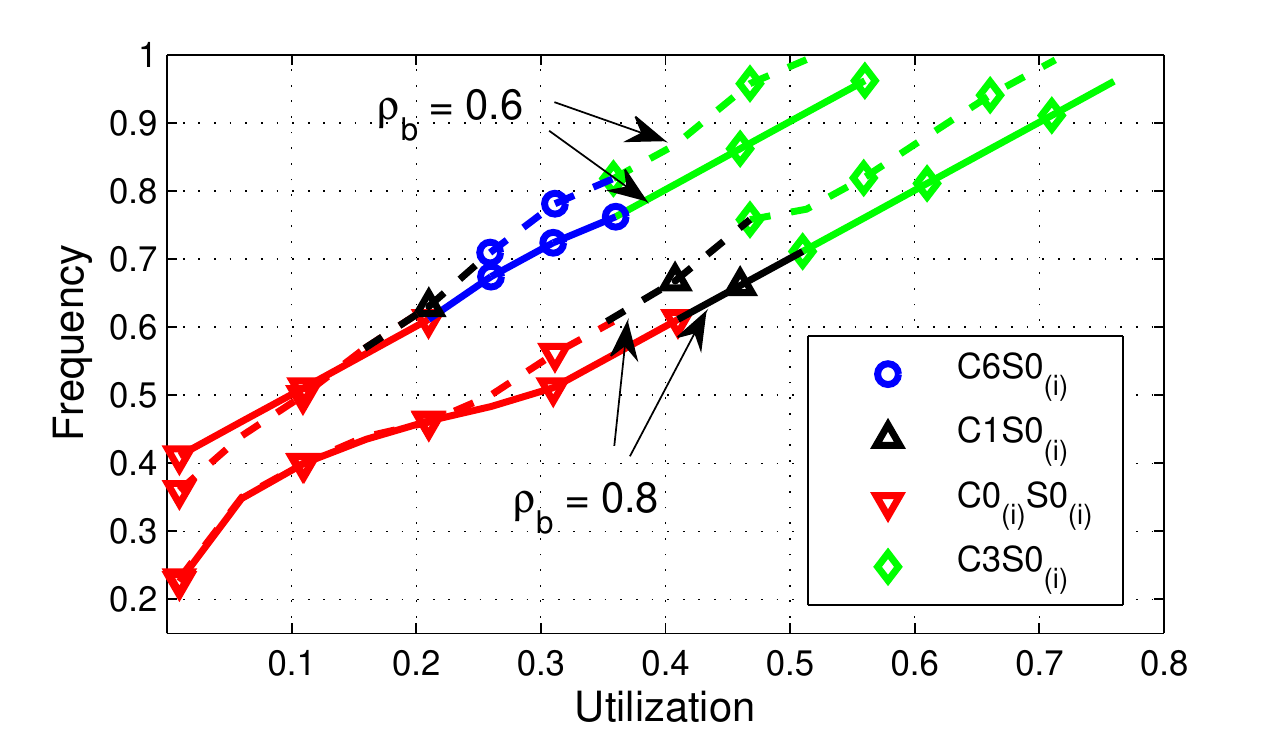}} 

\caption{Policy selection for DNS and Google-like workloads. Each
curve plots the optimal pairing of frequency setting and low-power
states as a function of utilization $\rho$, and parameterized by QoS
constraint $\rho_p$.  The different markers (see legend) indicate
which low-power state is optimal at each utilization. The solid lines represent what an idealized model computes and dashed lines represent the results using real-world workload statistics. 
} 

\label{fig.PP}
\end{figure*}


\subsubsection{Results and observations.}


We now present the results of our policy characterization according to
the workload statistics used by the BigHouse~\cite{BigHouse}
simulator. The BigHouse simulator stores statistics of
inter-arrival and service times accumulated from long-term
observation of live traffic traces for various real-world workloads.
Table~\ref{tb.applications} lists summary statistics (inter-arrival
and service time mean and coefficient of variation ($C_v)$) of three
workloads in the BigHouse simulator.  

Figure~\ref{fig.PP} shows the optimum policy as a function of workload, utilization,
and QoS constraint ($\rho_b$).  As an example, Figure~\ref{fig.PP}-(a)
plots the optimum policy for the DNS-like workload.  Operating
frequency is indicated on the vertical axis, utilization on the
horizontal axis, and the optimum choice of low-power state by the hash marks is indicated in
the legend. Depending on the utilization two different policies become
optimal. At low utilization $C0_{(i)}S0_{(i)}$ is optimal and at high
utilization $C6S0_{(i)}$ is optimal.  Two pairs of curves are
plotted.  The upper two curves are plotted for a baseline QoS
constraint set by $\rho_b = 0.6$, the bottom two for $\rho_b = 0.8$. Note that the constraint under $\rho_b = 0.6$
is {\em tighter} than the one under $0.8$.
In each pair of curves one is dashed, meaning it
is the policy choice based on the statistics of the BigHouse simulator
for that particular workload.  The other curve is solid, meaning that it is the optimum policy choice computed
by the model considered in Section~\ref{sec.singleserver} (i.e., Poisson job arrivals and exponential service times) with the
same mean inter-arrival and service time as its paired
BigHouse curve.


The plots in the second row of Figure~\ref{fig.PP} are defined
analogously except that the QoS is measured in terms
of the $95$th percentile response time, rather than in terms of
normalized mean response time.  Each sub-plot in the second row shares
the same workload as the plot directly above.  We summarize key
observations below.


 {\bf $1$) There is no ``one-size-fits-all'' policy.}  Different
workloads and different utilizations require different
policies. Almost all low-power states are useful for some set of
operating conditions.  Thus, relying on a single low-power state when
designing power-efficient architectures can be a poor choice.

{\bf $2$) The idealized model is sometimes good, but often 
one needs to use more realistic models.}  Recall that the model of
Section~\ref{sec.singleserver} is limited to idealized Poisson job arrivals and exponential service times.  These are analytically tractable
distributions and when they closely match the actual workload
statistics policies based on those results can perform almost as well
as the policies based on actual statistics.

We also note that the discrepancy between the policy results based on
the idealized model and the BigHouse model is different for two performance constraints; 
cf.~Figures~\ref{fig.PP}-(c) and -(d).  This is due to the fact that while the mean response time is only concerned with the mean, the $95$th percentile response time constraint is concerned with the tail behavior of the
response time distribution, which depends critically on the
variation in job size and inter-arrival time.  

{\bf $3$) Often the idealized model computes the best choice of
low-power state, but not the frequency setting.} Often for a given
utilization the computed optimal low-power state is the same but the frequency setting computed by the idealized model is lower than the one computed by BigHouse. If there is a way to adjust the frequency in runtime, one can rely simply on the idealized model without simulation to compute the optimal policy. We leave this as a part of our future work.

 {\bf $4$) The bump in low utilization region indicates that the
policy is exceeding its QoS constraint.}  At low utilization, the
optimal frequency curve for $C0_{(i)}S0_{(i)}$ often has a concave
shape. The consistency of this shape can be explained by referring
back to Figure~\ref{fig.Google_ER_different_rho}.  As was observed
earlier, at low utilizations the QoS constraint can be exceeded while
reaching the global minimum power (optimized across all frequencies).
In terms of Figure~\ref{fig.Google_ER_different_rho} this means that
the systems is operating strictly to the left of the vertical bar.
Since the same model underlies the policy optimization based on the
idealized and BigHouse models, the global power minimum will be the
same for both.  For this reason the BigHouse and idealized curves can
overlap at low enough utilizations.

As the utilization increases, the optimal frequency setting also
increases to keep the power at the global minimum.  However beyond
some utilization level (roughly $\rho = 0.3$ in
Figure~\ref{fig.Google_ER_different_rho}) the lowest power will no
longer continue to meet the target performance constraint.  At that
level the frequency needs to increase more quickly to continue to meet
the constraint.  Above that utilization levels the $C0_{(i)}S0_{(i)}$
curves in Figure~\ref{fig.PP} transition into their respective linear
regimes.  Since the $\rho_b = 0.6$ constraint is {\em tighter}
than the $\rho_b = 0.8$ constraint and so in curves with arrow ``$\rho_b = 0.6$'' in Figure~\ref{fig.PP} we
see no evidence of the bump.

\subsection{Runtime predictor}
\label{sec.runTimePredictor}

The runtime predictor works epoch by epoch, predicting for the current
epoch two important aspects of the workload based on its history: $1)$
inter-arrival time and service time statistics and $2)$
utilization. Each epoch is a $T$ minutes long ($T \geq 1$) time period.  The
prediction is fed into the policy manager.  The policy is updated at
the beginning of each epoch and is held constant throughout the epoch.

\begin{algorithm}[!t]
\caption{Pseudo-code for utilization predictor}
\label{alg.alg}
\begin{algorithmic}[1]
\STATE Initialize history depth $hist$.
\STATE Initialize a weight vector of size $p = hist$: $\mathbf{v} = \{v(1), \ldots v(p) \}$.  Set each of the $p$ entries to $1/p$.
\WHILE {prediction for $\rho(t)$}
\STATE \COMMENT{// Predict the utilization at time $t$, $\rho'(t)$ using LMS:}
\STATE Predict $\rho'(t) = \min [\sum_{i=1}^p v(i)  \rho(t-i),~1]$ from the past $p$ utilization values.
\STATE Compute $error = |\rho(t) - \rho'(t)|$.
\STATE Update weight $\mathbf{v}$ based on $error$ and $\rho (t-1 : t-p)$.
\IF {$error$ is larger than some adaptive threshold}
\STATE \COMMENT{// CUSUM test:}
\STATE Reset $p = 1$ and set $v(1) = sum(\mathbf{v})$.
\ELSE 
\STATE Grow $p$, $p = \min(p = p + 1,~hist)$ and set $v(i) = sum(\mathbf{v}) / p$ for all $1 \leq i \leq p$.
\ENDIF
\STATE $t = t + 1$
\ENDWHILE
\end{algorithmic}
\end{algorithm}

\subsubsection{Distribution prediction.}
\label{sec.distri_predict}
The inter-arrival time and service time distributions are predicted
based upon jobs events logged in previous epochs. The logs we collect
detail the arrival and service times of each jobs. These logged
statistics are scaled by the predicted utilization (to be discussed
later) and fed into the policy manager.  Using
Algorithm~\ref{alg.simulator} the policy manager then computes the
predicted power consumption and QoS of each candidate policy and then
selects the policy to use. Note that since we have already obtained the
workload log, generating jobs by sampling the distribution (step 1 in
Algorithm~\ref{alg.simulator}) is not needed. Logging all job events
is not necessary either: average behavior from the past several epochs
will suffice. 

Implicitly the predictor predicts the inter-arrival time and service
time distributions based on the past epochs. The motivation for
working with the logged arrival and service times is that
constructing, maintaining and updating a fine-grained distribution
histogram and simulation via sampling is expensive in both time and
space. Thus we find it is effective to use logs from previous epochs
without explicitly building a distribution. As shown in
Section~\ref{sec.methodology}, it takes only $6.3~ms$ for evaluating
one single policy. The policy manager only needs to determine the best
policy {\em once} every epoch.  Since, e.g., in the results of
Section~\ref{sec.evaluation} epochs will be in minutes-length and the determination of the best policy takes less than $1~s$
to compute, the computational overhead is  negligible.

\subsubsection{Utilization prediction.}
\label{sec.workload_est}

The distribution predictor predicts the statistics based on the recent epochs. We further enhance such prediction by a fine-grained, minute-by-minute utilization prediction. The workload log gathered in Section~\ref{sec.distri_predict} used to simulate the policies is adjusted based on the predicted utilization of the upcoming epoch (the first minute of the epoch, by default): the empirical inter-arrival times between jobs are scaled to match the upcoming predicted utilization.

\begin{figure}
\centering
\includegraphics[width = 0.5\textwidth]{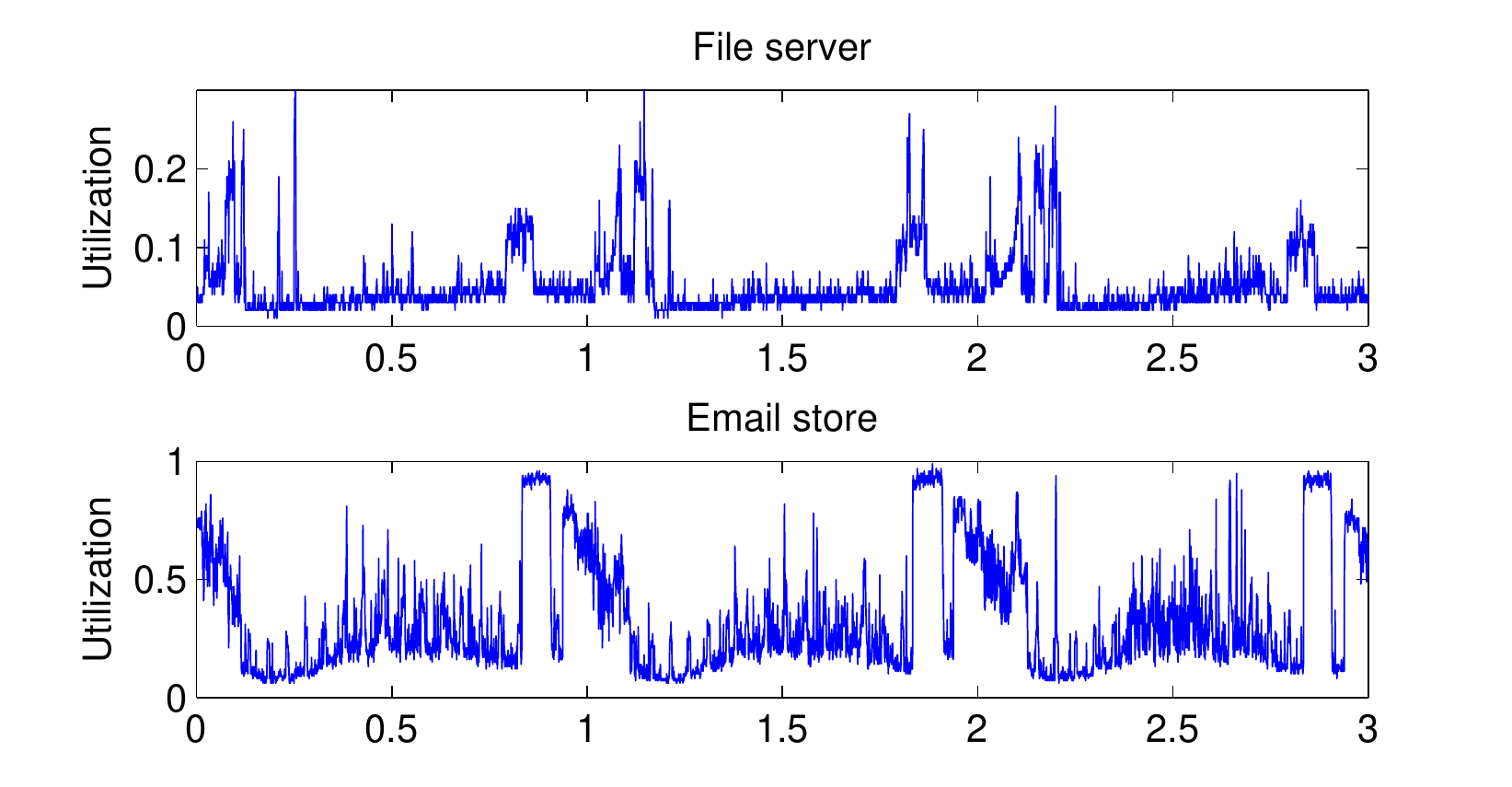}
\caption{Utilization traces plotted across 3 days for different services, both start at $12$ AM of a day. Email store is the host for student, faculty and staff email storage. File server is the host for student files \cite{WongAnnavaram, MedanBuyuktosunoglu}.}
\label{fig.workload}

\end{figure}

There is a large body of literature on utilization prediction;
much is based on pattern matching~\cite{GmachRolia} across days.  As
an illustration, in Figure~\ref{fig.workload} we plot several days
worth of minute-granularity utilization traces from academic
departmental servers.
We observe a periodic daily pattern to the utilization. The abrupt
surges observed towards the end of each day in the email store
workload, are due to maintenance and back-up services. In contrast to the earlier work based on pattern matching, we study fine-grained
minute-by-minute fluctuations in workload behavior. Our approach lets
the policy manager react at the processor level to real-time
fluctuations in the workload. Our examination will
reveal some fundamental aspects of what constitutes good prediction
and how the predictor should interact with the selected policy.

In SleepScale we implement three different utilization predictors: a naive-previous predictor, a least-mean-square adaptive filter (LMS)~\cite{Gustafsson}, and an LMS filter in conjunction with a cumulative sum change point detection (LMS+CUSUM)~\cite{Page}.

The naive-previous predictor simply uses the utilization in the last minute of the past $T$-minute epoch as the prediction for the current epoch. This predictor is best suited to track sudden changes in utilization, however it does not effectively predict the stationary behavior of the workload.

The LMS adaptive filter predicts the utilization based on a weighted combination of the utilizations observed over the past $p$
minutes.  The weights are updated every minute based on the prediction
error. The LMS adaptive filter outperforms the moving average
predictor (which would take the average utilization over the past $p$
minutes) because the weight for each of the past $p$ minutes is chosen
adaptively, rather than being fixed to a constant $1/p$. However, like the moving average predictor the LMS filter
smoothes the data, it does not track abrupt changes well. 

As an intermediary between naive-previous predictor and LMS filter, LMS+CUSUM does both tracking and stationary behavior prediction. The
pseudo-code of this LMS+CUSUM is given in the
Algorithm~\ref{alg.alg} box. When the
CUSUM algorithm detects an abrupt change, the look-back period $p$ in the LMS is
reset to $1$ (cf.~line~10).  This resetting drops the smoothing effect
of LMS and allows the filter to track the change better.  As long as
no further abrupt change is detected, $p$ grows until some maximum
value is reached (cf.~line~12).

\subsubsection{Dynamic frequency over-provisioning.} 
\label{sec.overprovision}

The utilization predicted for the first minute of the upcoming $T$-minute epoch is used to scale the workload log for the entire epoch. The larger $T$ is, the less likely the prediction
will be a good one for the entire $T$-minute epoch. If the
predictor overestimates the utilization realized in the epoch, jobs
will be processed faster and the queue will tend to empty. However, if
the predictor underestimates the realized utilization, the queue will
back up, and large delays may result, delays that can propagate into
subsequent planning epochs.  
To control for this in SleepScale we implement the following over-provisioning mechanism.

At the beginning of every $T$-minute epoch SleepScale computes the
average delay incurred by the jobs that were completed in the epoch
just past. If the average delay is {\em below} the delay in the
baseline system with utilization $\rho_b$, i.e., if it is less than
$\frac{1}{(1-\rho_b) \mu}$, then the frequency determined by the policy manager is further increased by a factor of $\alpha$. At first
glance this strategy may appear counter-intuitive since one might
think it is more natural to over-provision when the past delay has
been above (rather than below) the average.  However, we observe that such
over-provisioning works best as a ``guard band'' to buffer against
sudden increases in utilization. We illustrate the benefits and the costs of
over-provisioning in the next section.

%% file: evaluation.tex
To evaluate \pProfile~we combine the real-life daily utilization
traces discussed in Figure~\ref{fig.workload} with the BigHouse. We first generate sequences of jobs by sampling the inter-arrival time and service time cumulative distribution functions (CDF) from BigHouse~\cite{BigHouse}. In systems that serve only a single type of
job, the service time distribution is stationary.  What varies with
utilization is the distribution of inter-arrival times. In our simulated workload traces
we then scale the inter-arrival time between generated jobs to match the
time-varying utilization of Figure~\ref{fig.workload}. SleepScale uses the job stream as the causal input.

\subsection{Under real-world utilization}
\label{sec.underRealworldUtilization}
We first consider a DNS-like server following the email store
utilization trace of Figure~\ref{fig.workload}.  Across the day
the utilization trace covers a large range: from $0.1$
to $0.9$. We evaluate SleepScale over the period
extending from $2$ AM to $8$ PM as from $8$ PM to $2$ AM everyday
back-up and maintenance operations are scheduled. We select the baseline system
to be $\rho_b = 0.8$ resulting in a normalized mean response time
budget of $1/(1-\rho_b) = 5$. 

{\bf $1$) Utilization predictors and policy update interval.}
\begin{figure}
\centering
\includegraphics[width = 0.5\textwidth]{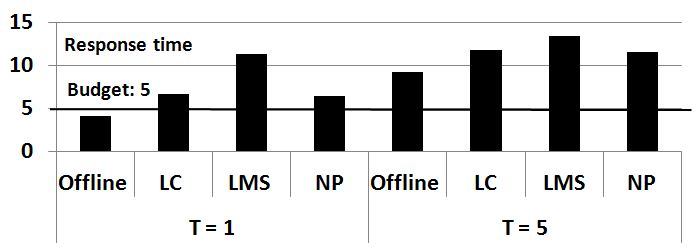}
\caption{Average response time under different predictors and policy update intervals. No over-provisioning is used (i.e., $\alpha = 0$).}
\label{fig.estimatorPolicyUpdate}
\end{figure}
In Figure~\ref{fig.estimatorPolicyUpdate} we study the performance of different predictors: LMS+CUSUM (LC), LMS-only (LMS), naive-previous (NP) and offline (Offline) as well as
the policy update periods ($T$). The offline predictor is a genie-aided predictor
where the true utilizations are assumed to be known non-causally in
advance. The naive-previous predictor simply uses the utilization in the last minute of the past $T$-minute epoch as the prediction for the duration of the next $T$-minute epoch. The LMS+CUSUM and LMS-only predictors are designed with history length $p = 10$. The average response time is measured when running SleepScale with no over-provisioning ($\alpha = 0$). 

We note that the more often SleepScale updates its policy (i.e., the smaller the $T$), the smaller the response time. Since SleepScale selects the best policy based on the predicted utilization, updating the policy fast often helps to mitigate the prediction error. We also note that the LMS+CUSUM predictor outperforms the LMS-only predictor since the former can track sudden changes in utilization.  However, in some cases this tracking can be detrimental: if a big surge is preceded by a short dip, the tracking
predictor may under-estimate the surge since it over-reacts to the
dip. For most of the utilization traces that we have, we notice that the accuracy of the naive-previous predictor is often comparable to that of the LMS+CUSUM predictor. The accuracy of these predictors can be further improved by considering the correlation (i.e., repeated daily patterns) across past days.

Finally we note that in Figure~\ref{fig.estimatorPolicyUpdate} the average response time exceeds the allowed budget in all cases when a utilization predictor is used. As mentioned in Section~\ref{sec.overprovision}, if
the predictor underestimates the realized utilization, the queue will
back up, and large delays may result, delays that can propagate into subsequent planning epochs.  Thus, we next set the over-provisioning factor $\alpha=0.35$ and compare the results to the other power management strategies. Recall from
Section~\ref{sec.overprovision} that this means the policy manager will
increase the frequency by $35\%$ if the delay in the
past $T$-minute epoch is within budget. 

\begin{figure}
\centering
\subfloat[Response time comparison]{\label{fig.avgDelay} \includegraphics[width=1\linewidth]{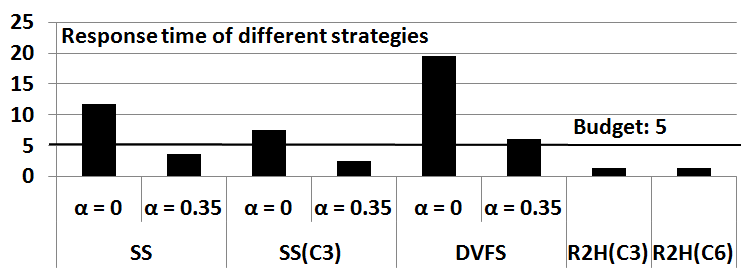}}  
 \\
\vspace*{-1 em}
\subfloat[Average power comparison]{\label{fig.avgDelay} \includegraphics[width=1\linewidth]{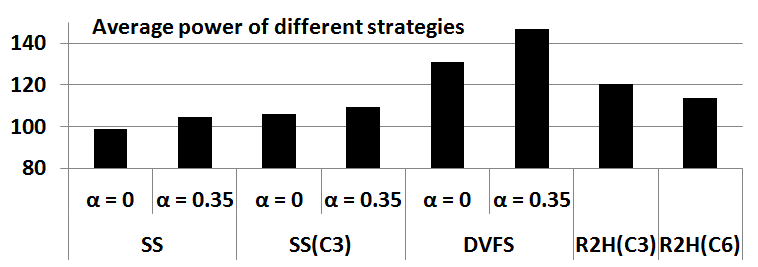}}  

\caption{Comparing SleepScale with other power control strategies. All strategies are running with LMS+CUSUM predictor with $p = 10$ and are updated every $T = 5$ minutes.}
\label{fig.policyComparison}
\end{figure}

{\bf $2$) Comparing \pProfile~with other strategies.}
In Figure~\ref{fig.policyComparison} we compare SleepScale (SS) with other power control strategies, including a SleepScale method that uses only low-power state $C3S0_{(i)}$ (SS(C3)), a DVFS-only strategy that only uses DVFS and no low-power state (DVFS) and race-to-halt mechanisms using $C3S0_{(i)}$ and $C6S0_{(i)}$ (R2H(C3) and R2H(C6) respectively). Both R2H(C3) and R2H(C6) correspond to the strategy that always operates at the
maximum frequency setting ($f=1$) and transitions into a low power
state ($C3S0_{(i)}$ or $C6S0_{(i)}$) immediately upon the queue
emptying. All strategies use the LMS+CUSUM predictor with the history length $p = 10$ and are updated every $T = 5$ minutes. We make a number of observations. 

SleepScale, when equipped with the frequency over-provisioning, achieves the best power efficiency of all strategies while maintaining the response time within budget. Among others, using DVFS only wastes power as the server is not allowed to enter any low-power state when idling. The race-to-halt mechanism also consumes more power than SleepScale as it sets frequency to the maximum. When SleepScale is set to use only $C3S0_{(i)}$ it also consumes more power as this low-power state is a sub-optimal one. {\em Our results clearly demonstrate the importance of joint optimization of speed scaling and sleep state selection.}

The properly set over-provisioning reduces response time at the cost of a slight increase in power. This is due to the fact that the extra
capacity over-provisioning allocates during periods of low utilization allows the server to accommodate unpredictable surges in utilization. This proves to be essential to meet the mean response time budget. Also running slightly faster does not cost too much power as the sever can enter low-power state sooner. (This is not true for DVFS-only strategy as it has no low-power state to use.)

Finally we comment on the large response time in the DVFS-only strategy. To minimize power under a given response time budget, the optimal solution for the DVFS-only strategy is to set the frequency low enough that the response time just meets the budget. This however consumes all the performance budget thus even a slight utilization miss-prediction can result in large queuing delay. However for SleepScale, as we demonstrate in Figure~\ref{fig.Google_ER_different_rho}, does not always consume the entire performance budget.

\subsection{Distribution of low-power states}
\begin{figure}
\centering
\includegraphics[width = 0.5\textwidth]{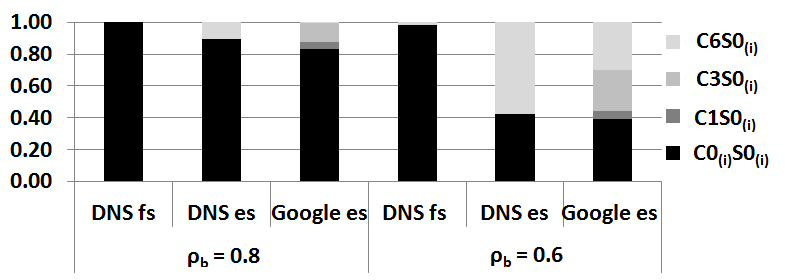}
\caption{Distribution of optimal low-power states selected by SleepScale. LMS+CUSUM predictor is used with history length $p = 10$, update interval $T = 5$ and over-provisioning $\alpha = 0.35$.}
\label{fig.policy_distri}
\end{figure}

In Figure~\ref{fig.policy_distri} we present the distributions of the low-power states selected by \pProfile~for different workloads,
baselines, and utilizations. We report results for the file server
(fs)  and email store (es) utilizations. We run both DNS
and Google-like services with $\rho_b = 0.6$ and $\rho_b =
0.8$. 

At low average utilization and when there are few time-varying
fluctuations (which is the case for file server), a single low-power
state often suffices. For highly time-varying utilization traces (such
as email store), multiple low-power states are used: $C0_{(i)}S0_{(i)}$ and $C6S0_{(i)}$. Tightening
the performance constraint further will lead to deeper sleep states
being used more often, as the fast processing required to meet the
budget creates more opportunities for entering aggressive
power-saving states like $C6S0_{(i)}$. These observations all
match our analyses in
Section~\ref{sec.singleserver}.

%% file: conclusion.tex

In this paper we develop \pProfile, a power management tool to manage
data centers for power efficiency while meeting QoS
agreements.  \pProfile~uses queuing-based simulations to
determine how best to exploit the low-power states and operating frequency built
into modern CPUs.  The optimum such policy can be
determined at runtime, via online prediction for workload statistics and utilization. We characterize the
performance of \pProfile~on data center traces.  We evaluate
\pProfile~realizing significant power savings relative to some
conventional power management strategies while meeting the same
QoS constraints.

More power control knobs are being built into different hardwares in computer systems. However challenges remain in effectively adjusting these knobs in a coordinated manner. One future research direction involves not only the power control mechanisms in CPUs, but also the ones in other system components. We conjecture that high-level queuing models and simulations (rather than fine-grained instruction-level simulation) can help in designing effective power control strategies. Another research direction involves studying SleepScale on multi-core, multi-server systems in order to scale out. The focus is on controlling the overall queuing simulation overhead, although SleepScale can be performed on each core or server independently.

%% file: ack.tex
The authors would like to thank Aman Chadha (UW-Madison) for early efforts on workload generation, Daniel Wong (USC) for the departmental datacenter utilization traces, Daniel Myres (UW-Madison) for the discussion on queuing theory, Ken Vu, Srini Ramani (IBM) for their continuous suggestion and support, and the anonymous reviewers for their feedback. 

This work is
supported in part by NSF grants (CCF-0963834, CCF-0953603, and CNS-1217102) and an AFOSR grant (FA9550-13-1-0138). Nam Sung Kim has a financial interest in AMD.

%% file: theorems.tex
\section*{Appendix}
\label{sec.singleServerThms}
The analytic results extend the work in \cite{yanpeiStarkNamCISS}. The average power consumption for the single-server system
with $n$ low-power states described in Section~\ref{sec.singleserver} is
\begin{equation*}
\label{eq.singleserverpower}
\bbE[P] \! = \! \frac{1}{\lambda L} \! \left [ \! \sum_{i=1}^{n-1} P_i (e^{-\lambda \tau_i} \!-\! e^{-\lambda \tau_{i+1}} ) + P_n e^{-\lambda \tau_n} \! \right] \! +  \! P_0 \!\left(\!\! 1 - \frac{e^{-\lambda \tau_1}}{\lambda L} \!\! \right )
\end{equation*}
where $L$ is defined as
\begin{align*}
L = \frac{ \mu f  + \mu f \lambda \left[\sum_{i=1}^{n-1} w_i (e^{-\lambda \tau_i} - e^{-\lambda \tau_{i+1}} ) + w_n e^{-\lambda \tau_n} \right ]}{\lambda(\mu f - \lambda)}.
\end{align*}
This can be proved using busy period analysis and first
principles, see~\cite{mor_book}. The mean response time $\bbE[R]$ is
\begin{align*}
\label{eq.xxx}
\bbE[R] = \frac{1}{\mu f - \lambda} + \frac{2 \bbE[D] + \lambda \bbE[D^2]}{2 ( 1 + \lambda \bbE[D] )},
\end{align*}
where 
\begin{align*}
\bbE[D^{\alpha}] = \sum_{i=1}^{n-1} w_i^{\alpha} (e^{-\lambda \tau_i} - e^{-\lambda \tau_{i+1}} ) + w_n^{\alpha} e^{-\lambda \tau_n}.
\end{align*}
This can be derived using the
result from \cite{Welch} and some algebraic manipulations. Both $\bbE[R]$ and $\bbE[P]$ can be extended to the case where service time is not exponential.

The probability the response time $R$ exceeds deadline $d$ is
\begin{align*}
Pr(R \geq d) = \frac{e^{-(\mu f - \lambda) d} - w_1 (\mu f - \lambda) e^{- d /w_1}}{1 - w_1(\mu f - \lambda)}. 
\end{align*}
Note that when
$d = 0$ the $Pr(R \geq d) = 1$.  When $d = \infty$ the $Pr(R \geq d) = 0$.  When $w_1 = 0$ the $Pr(R \geq d) = e^{-(\mu f - \lambda) d}$. When $w_1 = \infty$ the $Pr(R \geq d) =
1$. This result can be proved using a Laplace transform
analysis of the response time $R$.